\documentclass{iaus}
\usepackage{graphicx}

\newcommand{\Msun}{M$_{\odot}$}

\title[Z interpolations] 
{Are interpolations in metallicity reliable?}

\author[Angeretti et al.]   
{L.Angeretti$^1$ G. Fiorentino$^1$ \break \and L. Greggio$^2$}

\affiliation{$^1$ INAF-OA Bologna, via Ranzani, 1 I-40127, Bologna, Italy,
\break email: luca.angeretti@oabo.inaf.it, giuliana.fiorentino@oabo.inaf.it.
 \\[\affilskip]
$^2$ INAF-OA Padova, vicolo dell'Osservatorio 1, Padova, Italy
\break email: laura.greggio@oapd.inaf.it}

\pubyear{2004}
\volume{241}  
\pagerange{119--126}
\date{?? and in revised form ??}
\jname{Proceedings Title IAU Symposium}
\editors{A.C. Editor, B.D. Editor \& C.E. Editor, eds.}
\begin{document}

\maketitle

\begin{abstract}
In this proceeding we present the procedure that we have adopted to
obtain a dataset of Padova94 tracks (\cite{bre93,fag94a,fag94b})
interpolated in metallicity. The procedure requires special care to
avoid spurious features in the resulting grid, thus we have subdivided
tracks in evolutionary phases, we have chosen the suitable
interpolation method and the transition masses. Finally, we have
compared our interpolated dataset with a similar models, \cite{gir00},
obtaining a general good agreement.
\end{abstract}

\firstsection 

\section{Introduction}
The Star Formation History (SFH) of stellar populations (e.g. galaxies
and clusters) may be derived using the Synthetic CMD Method (SCM)
based on the interpolation in mass and age of the Stellar Evolutionary
Models (SEM).

To perform systematic studies and  comparisons of the SFH of galaxies,
the SEMs should  have: {\it i}) a dense grid over  a large masse range
(up to 100 \Msun) to cover the entire age of the Universe; {\it ii}) a
homogeneous input physics; {\it iii})  a dense grid in metallicity (Z)
to cover the large range in metallicity of the stellar populations.

At  present, only  a few  SEMs datasets  satisfy two  of  the previous
points: Padova94 (\cite{bre93,fag94a,fag94b})  dataset is one of them,
but its grid in metallicity is too coarse, particularly in the Z range
0.0004 -  0.004, that  is the most  commonly used.  To  compensate for
this limitation,  we have interpolated  in this metallicity  range the
Padova94 models.

In this  work, we discuss (section  2) our cares  and prescriptions to
obtain the  interpolated tracks for  Z=0.001 (fig. 1).  The comparison
between  our results  and \cite[hereafter  G00]{gir00} models,  of the
same metallicity,  are shown in section  3, in order  to emphasize the
reliability of the interpolation.

\section{The method}
The  interpolation  in  metallicity  require  special  care  to  avoid
spurious features  in the resulting  grid. Our procedure  goes through
three steps:
\begin{enumerate}
  \item The subdivision of  the tracks in specific evolutionary phases
with: {\it\sf  i}) the same  internal burning processes;  {\it\sf ii})
comparable lifetimes;  {\it\sf iii}) similar behaviors on  the HRD. \\
The subphases differ for the  low, intermediate or high masses because
of the different structures  on the HRD.  Interpolations are performed
at constant fractionary age within each subphase.
  \item We  have performed several  tests within Padova94 sets  to get
the reliable  interpolation method. We  have found that an  average of
linear and logarithmic  factor in Z gives us the  best result. The age
has been linearly interpolated.
  \item We choose the transition masses for the selected metal content
taking into  account the trend of  the transition masses  over all the
metallicity range.
\end{enumerate}

\section{Comparison and Conclusions}
G00 models  are similar to Padova94 models,  although some differences
are  present  because  of   different  input  physics.   We  test  our
interpolated tracks by comparing them with G00 models at Z=0.001:
\begin{enumerate}
  \item The  $dt/t [= (t_{inter}  - t_{Gir})/ t_{inter}]$ of  the {\it
H-burning phase} is less than 2\% for M $>1.4$ \Msun. and smaller than
15\% for M $<1.4$ \Msun. The latter discrepancy reflects the different
definition  of the  overshooting  parameter in  the  mass range  1-1.4
\Msun.
  \item The $dt/t$ of the {\it He-burning phase} is less than 15\% for
M $<1.7$ \Msun \, and smaller  than 7\% for M $>1.7$ \Msun.  For M=1.7
\Msun \, the He-burning lifetime of our model is two times larger than
that of G00. This is because in  our set the M=1.7 \Msun \, is adopted
to be  below the transition mass, while  in G00 models it  is close to
the transition mass.
\end{enumerate}

Taking into  account the differences in  the input physics,  we find a
good general  agreement by comparing the interpolated  tracks with the
G00 models (e.g. the lifetime  of the major phases are reasonably well
evaluated)  although some  difference  are present  especially in  the
extention and luminosity of the blue loops.


The results  showed here refers to  a specific choice of  Z.  The same
interpolation  algorithm   has  been  succesfully   applied  to  other
interpolated dataset (Z=0.002,0.003)  and for other metallicity range
(Z=0.004-0.008).

\begin{figure}
 \includegraphics[height=6truecm]{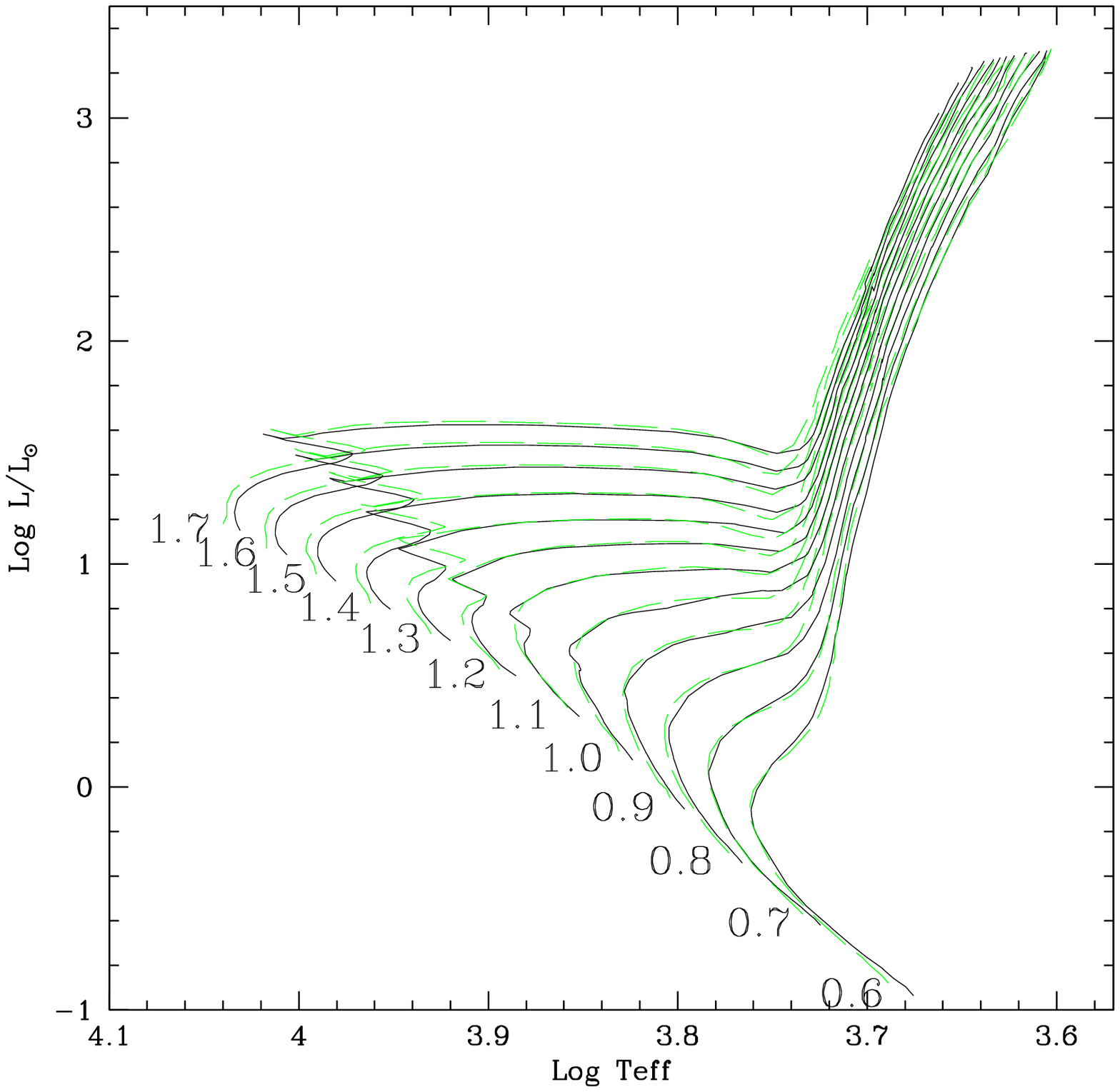}
 \includegraphics[height=6truecm]{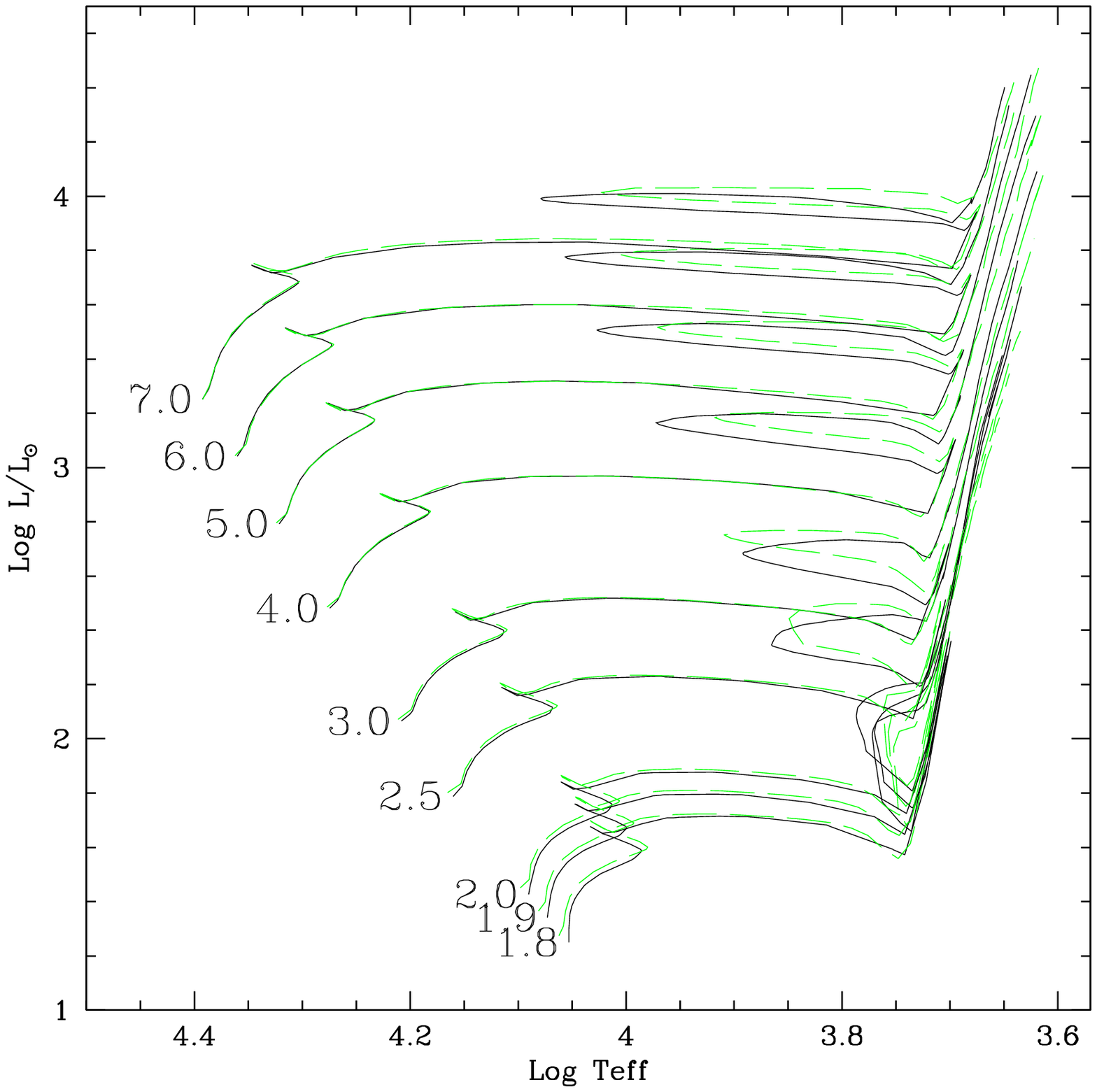}
  \caption{ Our interpolated models for Z=0.001 (dashed lines) and G00
models (solid lines) of the same metallicity, for low (left panel) and
intermediate (right panel) masses tracks.}\label{fig1}
\end{figure}

\end{document}